\documentclass[%
 reprint,
 superscriptaddress,
 amsmath,amssymb,amsfonts,
 aps,
 prl,
 floatfix,
 longbibliography
]{revtex4-2}

\usepackage[utf8]{inputenc}
\usepackage[pdftex]{graphicx} \graphicspath{{}}
\usepackage{float} \usepackage{color}
\usepackage{comment}
\usepackage[pdftex,colorlinks=true]{hyperref}
\usepackage{subfigure}
\hypersetup{
  colorlinks=true,
  linkcolor=blue,
  citecolor = blue,
  urlcolor=blue,
}
\usepackage{physics} 

\usepackage{mathtools}
\DeclarePairedDelimiterX\proj[2]{\delimsize\vert#1\rangle}{\langle#2\delimsize\vert}{ }
\usepackage{siunitx}

\usepackage{tikz}
\newcommand*\emptycirc[1][0.4ex]{\tikz\draw (0,0) circle (#1);} 
\newcommand*\fullcirc[1][0.4ex]{\tikz\fill (0,0) circle (#1);}

\begin{document}

\title{Emergent interaction-induced topology in Bose-Hubbard ladders}

\author{David Wellnitz}
\affiliation{JILA, National Institute of Standards and Technology and Department of Physics, University of Colorado, Boulder, CO, 80309, USA}
\affiliation{Center for Theory of Quantum Matter, University of Colorado, Boulder, CO, 80309, USA}
\author{Gustavo A. Dom\'inguez-Castro}
\affiliation{Institut f\"ur Theoretische Physik, Leibniz Universit\"at Hannover, Appelstrasse 2, D-30167 Hannover, Germany}
\author{Thomas Bilitewski}
\affiliation{Department of Physics, Oklahoma State University, Stillwater, Oklahoma 74078, USA}
\author{Monika Aidelsburger}
\affiliation{
Max-Planck-Institut f\"ur Quantenoptik, Hans-Kopfermann-Strasse 1, 85748 Garching, Germany}
\affiliation{Fakult\"at f\"ur Physik, Ludwig-Maximilians-Universit\"at, Schellingstrasse 4, 80799 M\"unchen, Germany}
\affiliation{Munich Center for Quantum Science and Technology (MCQST), Schellingstraße 4, 80799 M\"unchen, Germany}
\author{Ana Maria Rey}
\affiliation{JILA, National Institute of Standards and Technology and Department of Physics, University of Colorado, Boulder, CO, 80309, USA}
\affiliation{Center for Theory of Quantum Matter, University of Colorado, Boulder, CO, 80309, USA}
\author{Luis Santos}
\affiliation{Institut f\"ur Theoretische Physik, Leibniz Universit\"at Hannover, Appelstrasse 2, D-30167 Hannover, Germany}

\date{\today}

\begin{abstract}
We investigate the quantum many-body dynamics of  bosonic atoms hopping in a two-leg ladder with strong on-site contact interactions. We observe that when the atoms are prepared in a staggered pattern with pairs of atoms on every other rung, singlon defects, i.e.~rungs with only one atom, can localize due to an emergent topological model, even though the underlying model in  the absence of interactions admits only topologically trivial states. This emergent  topological localization results from the formation of a zero-energy edge mode in an effective lattice formed by  two  adjacent   chains with alternating strong and weak hoping links (Su-Schrieffer-Heeger chains) and opposite staggering  which interface at the defect position.
Our findings open the opportunity to dynamically generate non-trivial topological behaviors without the need for complex Hamiltonian engineering.
\end{abstract}
\pacs{}

\maketitle



Topological phases of matter \cite{RevModPhys.89.041004,SPT_review_2015}, including the  quantum Hall effect~\cite{Prange1990,Nature_Review_QH_2020,RevModPhys_QAH_23}
and topological insulators~\cite{Hasan2010,RevMod_Phys_Cheng_2011}   
have  attracted significant attention for quantum science and technologies \cite{Luo2022,Gilbert2021,Breunig_2021,RevModPhys.80.1083} given 
their robustness against disorder and defects. While  non-interacting topological phases have been realized in  a broad range of settings, both classical \cite{PhysRevX.5.021031,Hafezi_2013,St_Jean_2017,S_sstrunk_2015} and quantum \cite{Atala2013,Atala2014,Stuhl2015,Mancini2015,Jotzu2014,Aidelsburger2014,Lohse2015,Nakajima2016,Aidelsburger2018,Cooper2019}, and are to a great extent well understood \cite{Ludwig_2015,RevModPhys.88.035005},  interaction-enabled topological states remain largely unexplored. Understanding how topological phases may arise in interacting systems is hence an exciting, but challenging problem.


\begin{figure}[t!]
\centering
\includegraphics[width=0.85\columnwidth]{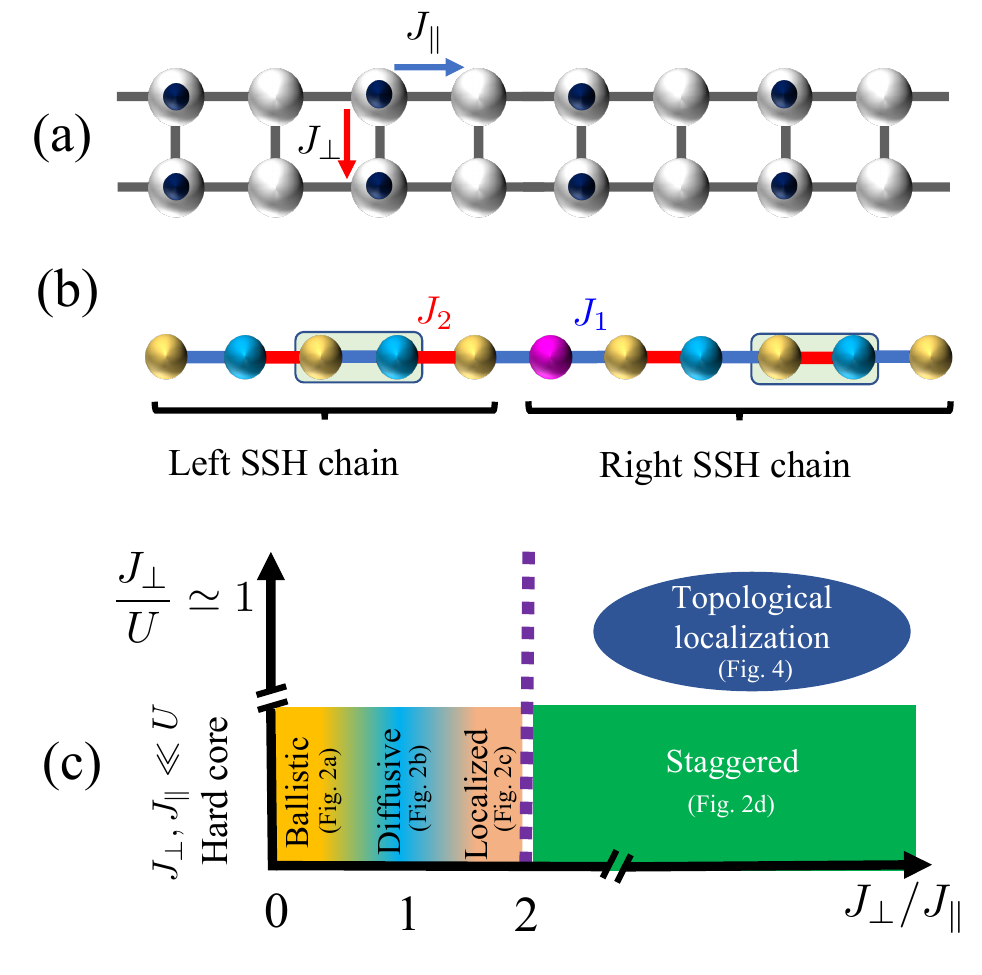}
\vspace{-0.4cm}
\caption{
(a) Bose-Hubbard ladder with rung~(leg) hopping $J_\perp$~($J_\parallel$). Initially there is one~(no) particle per 
site in even~(odd) rungs. (b) Emergent SSH chains of opposite staggering interface at a central site~(purple), and emergent model in the dynamics of singlon defects. (c) Overview of the regimes. In the hard-core regime, $J_\perp, J_\parallel \ll U$, rung density-density (RDD) correlations are negative, and their expansion within relevant time scales transitions from ballistic to diffusive to localized when $\eta\equiv J_\perp/J_\parallel$ grows from $0$ to $2$. For $\eta>2$, staggered RDD correlations expand ballistically. For $J_\parallel \ll J_\perp \lesssim U$, singlon defects experience topological or non-topological localization. 
We indicate which figures illustrate the regimes.
}
\label{fig:1}
\end{figure}


Ultra-cold gases \cite{Bloch2008, Bloch2012,Gross2017,Schaefer_2020} with tunable interactions in optical lattices and tweezer arrays~\cite{Browaeys2020} are emerging as an excellent platform to shed light on this direction~\cite{Goldman2016, AiP_Top_atoms_2018,Cooper2019}, in particular with their capability to observe many-body states with single-site and spin resolution \cite{Bakr_2009,Nature_Sherson_2010,PRL_Cheuk_2015,Haller2015,Parsons2015, Edge2015,Boll_2016,Gross2021}. These have allowed the realization of non-interacting topological phases and observation of key underlying features~\cite{Atala2014,Stuhl2015,Mancini2015,Jotzu2014,Aidelsburger2014,Lohse2015,Nakajima2016}, such as edge states and currents \cite{Atala2014,Stuhl2015,Mancini2015,Braun_2024}, Chern numbers \cite{Aidelsburger2014}, and topological  pumping \cite{Lohse2015,Nakajima2016}. Progress towards the implementation of  interacting systems such as many-body symmetry protected topological phases \cite{SPT_review_2015} include bosonic Su-Schrieffer-Heeger (SSH) models~\cite{PhysRevLett.42.1698}  in interacting Rydberg atoms~\cite{ Leseluc2019}, 
a fractional quantum Hall state 
with few atoms
~\cite{Leonard2023,Lunt2024Realization}, 
and the Haldane phase in Fermi–Hubbard ladders~\cite{Haldane_83,Sompet2022}. However, especially for 
mobile particles, 
reaching the ultracold temperature necessary to observe interacting topological ground states remains a significant obstacle. While internal degrees of freedom can be 
pumped into a single state with essentially zero entropy, similar techniques do not exist for motion in lattices. It would be hence highly appealing to find settings 
where  topology emerges naturally in the dynamics of a strongly interacting system ~\cite{AnuRev_Langen_2015,Bloch2008, Bloch2012,Gross2017,Schaefer_2020,DAlessio2016} from an easy to prepare initial state.

In this Letter,  we report on emerging interaction-induced topological localization in an experimentally accessible system of   strongly interacting bosons in two-leg ladders. Two-leg ladders have played a major role as model systems for quantum magnetism~\cite{Dagotto1996,Dagotto1999}, in non-equilibrium many-body dynamics, hydrodynamics and transport~\cite{Steinigeweg2014, Schubert2021, Rakovszky2022,dominguez2023relaxation,Hirthe_2023}, 
and in realizations of synthetic magnetism~\cite{Atala2014}. More recently, bosons in optical ladders have been used to study non-equilibrium dynamics in the hard-core limit where strong interactions prevent 
more than one atom per site~(a model equivalent to an XX spin ladder~\cite{Steinigeweg2014}). By initially preparing  a density wave~(DW) along the legs with filled~(empty) sites in even~(odd) rungs~(Fig.~\ref{fig:1}~(a)), experiments have observed within their accessible time scales a crossover from ballistic to diffusive correlation dynamics 
when the rung hopping~($J_\perp$) increases from zero to equal to the leg hopping~($J_\parallel$)~\cite{wienand2023emergence}. 

Here  we analyze  the opposite  regime  when $\eta\equiv J_\perp/J_\parallel\gg 1$, focusing on experimentally accessible time scales.
Besides radically different 
correlations in the hard-core limit, we find that, surprisingly,  when the  on-site interactions are  large but finite,
an initial isolated singlon defect in the DW pattern, i.e. a singly-occupied rung, experiences an emergent effective lattice composed of two SSH chains of opposite staggering, and hence different topology, that meet at the initial defect position~(Fig.~\ref{fig:1}~(b)). 
These interfaces can feature two distinct types of localized states resulting in the localization of defects: (i) zero energy edge modes related to the SSH topology, and (ii) energetically bound states at the ends of the energy spectrum.
Our analysis hence unveils a surprising link between topology and many-body dynamics in strongly interacting ladders. Contrary to other realizations of topological models in optical lattices~\cite{Leder2016,Leseluc2019,Sompet2022}, or topological interfaces of SSH chains in other
systems~\cite{Poli2015, Meier2016, Cai2019, Lin2021}, the SSH chain and the topological interface are not externally implemented, but 
emerge naturally from the interplay of interactions, motion and the original DW pattern. This intriguing physics~(Fig.~\ref{fig:1}~(c)) can be probed in on-going experiments.



\paragraph{Model.--} We consider bosons in a square ladder, see Fig.~\ref{fig:1}~(a), with legs 1 and 2,  described in the tight-binding regime by the Bose-Hubbard~(BH) Hamiltonian $\hat H = \hat H_0 + \frac{U}{2} \sum_{i,\alpha}  \hat{n}_{i,\alpha} (\hat{n}_{i,\alpha} -1)$, with
\begin{equation}
\!\!\hat H_0 \! =\! -\! \sum_{i} 
\!\left(
\frac{J_\parallel}{2} \!\!\sum_{\alpha=1,2} 
\hat{b}^{\dagger}_{i+1,\alpha} \hat{b}_{i,\alpha} 
+ \frac{J_\perp}{2} \hat{b}^{\dagger}_{i,1} \hat{b}_{i,2} 
+ \mathrm{H. c.} \!
\right),
\label{eq:H0}
\end{equation}
where $\hat b_{i,\alpha}$ is the bosonic operator at site $i$ of leg $\alpha$, $\hat n_{i,\alpha}=\hat b_{i,\alpha}^\dag \hat b_{i,\alpha}$, and $U$ characterizes the on-site interactions. Motivated by recent experiments~\cite{wienand2023emergence}
we assume large $U$, and consider an initial DW, 
in which sites at even~(odd) rungs are occupied by one~(no) atom.



\begin{figure}
\centering
\includegraphics[width=0.85\columnwidth]{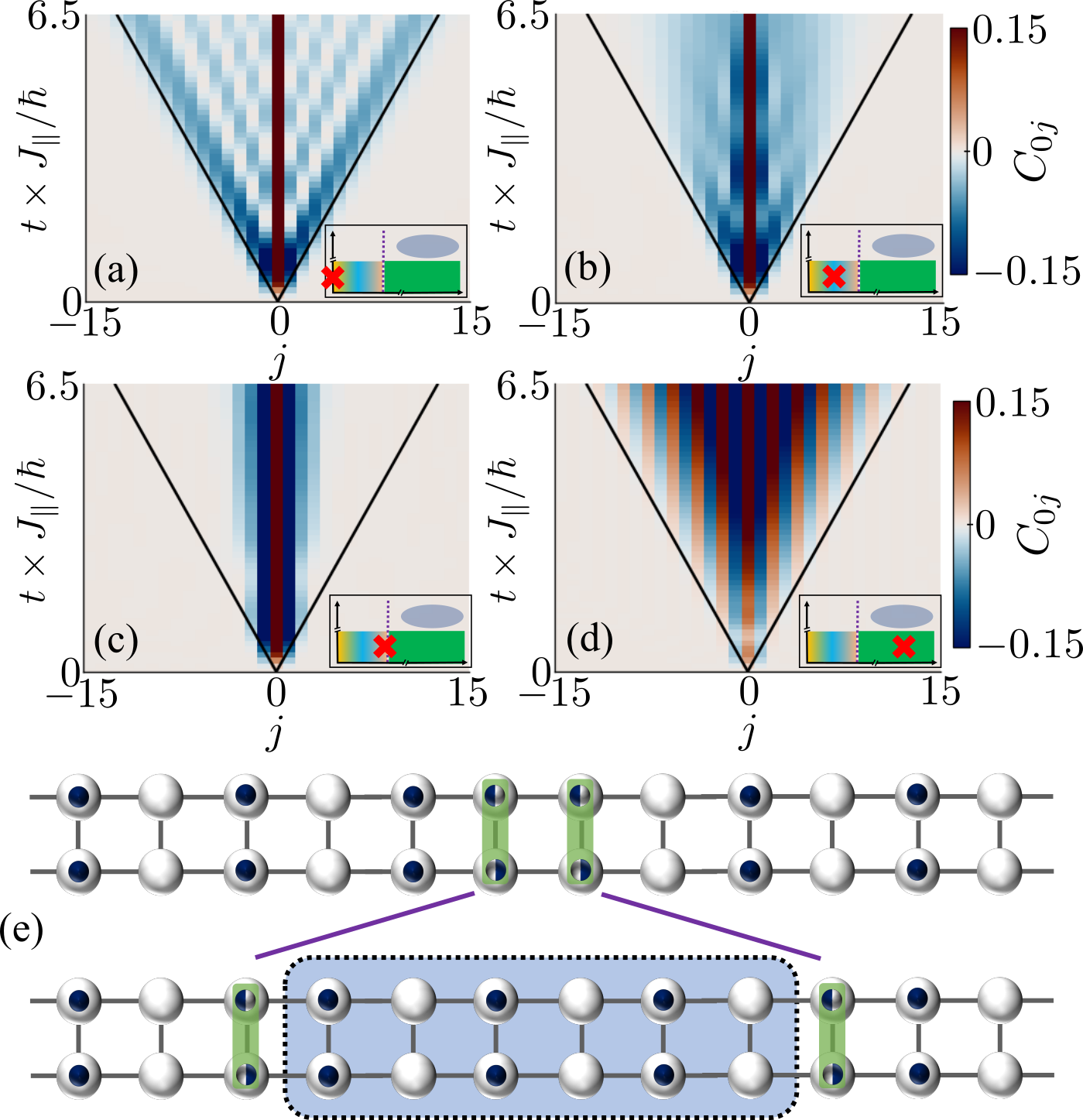}
\caption{Time evolution of the RDD correlation $C_{0j}$ with $\eta=0$~(a), $\eta=1$~(b), $\eta=2$~(c) and $\eta=8$~(d). Simulations of the hard-core Hamiltonian $\hat H_0$ on $31$ sites, using MPS
with bond dimensions $\chi = 1536$ (a/b), 1024 (c), 512 (d). The black lines correspond to the ballistic propagation when $\eta=0$. The red ``x'' in the insets locates the plots in Fig.~\ref{fig:1}(c). (e) Mechanism leading to the staggered correlations in (d). Quantum fluctuations result in pairs of singlon defects~(green), which when moving apart form a string of rungs~(blue shaded area) with an inverted DW compared to the initial one.
}
\label{fig:cones}
\end{figure}




\paragraph{Hard-core regime.--} 
To develop an intuition of the dynamical regimes, we first consider an idealized scenario without initial defects in the DW. Furthermore, we restrict the dynamics to the hard-core manifold described solely by $\hat H_0$ for very large $U \gg J_\perp,J_\parallel$.
A Bogoliubov stability analysis
shows that the DW is stable against the exponential proliferation of singlon defects if $\eta>2$~(see~\cite{SM}). Indeed, $\eta\simeq 2$ marks the onset of clearly different dynamics. As in recent experiments~\cite{wienand2023emergence}, we consider rung density-density (RDD) correlations 
$C_{ij} = \left< \hat{n}_i \hat{n}_j \right> - 
\left<\hat{n}_i \right> \left< \hat{n}_j\right>$, with 
$\hat{n}_i = \hat{n}_{i,1} + \hat{n}_{i,2}$ the particle number operator of rung $i$. Our results of $C_{0j}(t)$, obtained using Matrix Product State~(MPS) calculations, 
show a markedly different behavior for $\eta <2$ than for $\eta>2$. For $\eta=0$~(Fig.~\eqref{fig:cones}~(a)), the system is integrable, and the correlations expand ballistically as $C_{0j}(t)=-\frac{1}{4}{\cal J}_j^2\left(2J_\parallel t/\hbar\right)$, with ${\cal J}_j$ the Bessel function of first kind. As discussed in Ref.~\cite{wienand2023emergence}, when $\eta$ increases from $0$ to $1$ the expansion changes from ballistic to diffusive within the time scale of the experiments~(Fig.~\eqref{fig:cones}~(b)). A further increase of $\eta$ results in a strong slowdown of the evolution of the correlations, which within experimentally accessible time scales becomes clearly subdiffusive, and eventually  approximately localized at $\eta=2$~(Fig.~\ref{fig:cones}~(c)). One would naively expect that the larger $\eta$ the more localized $C_{0j}$ would be. Interestingly, the nature of the correlations changes remarkably at $\eta\simeq 2$. Whereas $C_{0j}<0$ for $\eta<2$, for $\eta>2$ RDD correlations are staggered, $(-1)^j C_{0j}>0$, and expand ballistically as for  $\eta=0$~(Fig.~\ref{fig:cones}~(d)).

To understand these dynamics for $\eta \gg 1$, we introduce the hard-core rung states:
$|2\rangle \equiv \begin{pmatrix}
\vspace{-0.2cm}
\fullcirc\\
\fullcirc
\end{pmatrix}$, 
$
|0\rangle \equiv \begin{pmatrix}
\vspace{-0.2cm}
\emptycirc\\
\emptycirc
\end{pmatrix}$, and 
$|\pm\rangle \equiv \frac{1}{\sqrt{2}}\left [\begin{pmatrix}
\vspace{-0.2cm}
\fullcirc\\
\emptycirc
\end{pmatrix} 
\pm
\begin{pmatrix}
\vspace{-0.2cm}
\emptycirc\\
\fullcirc
\end{pmatrix} 
\right ]$, 
where $\fullcirc$~($\emptycirc$) denotes an  occupied~(empty) site.
Quantum fluctuations due to leg hopping create singlon-defect pairs~(Fig.~\ref{fig:cones}~(e)) :
$|2,0\rangle\to |+,+\rangle - |-,-\rangle$, 
with a density $1/\eta^2$.
These defects, initially at neighboring sites, drift apart with rate $J_\parallel$ by position swaps between $|\pm\rangle$ and $|2\rangle$ or $|0\rangle$ induced by the leg hopping~(Fig.~\ref{fig:cones}~(e)). After a time $t$, the 
defects have a probability 
${\cal J}_{r-1}^2(2J_\parallel t)$ to be at $r\geq 1$ sites apart~\cite{SM}.
The rungs in between the defects present an inverted DW pattern compared to the original one~(Fig.~\ref{fig:cones}~(e)), and the RDD correlations acquire the form~\cite{SM}:
\begin{equation}
\!\!\!\!C_{0j}(t)\!\propto \! (-1)^j \!\left (
\!{\cal J}_{j-1}^2(2J_\parallel t)
\!+\! 
4\sum_{k>0}\! k{\cal J}_{k+j-1}^2(2J_\parallel t)\!\right ),
\end{equation}
which corresponds to the staggered correlations of Fig.~\ref{fig:cones}~(d), which expand ballistically as for $\eta=0$. 



\paragraph{Effective SSH chain.--} Up to this point we have considered the hard-core model~\eqref{eq:H0}. Large but finite $U$ may however play an important and surprising role in the defect dynamics. Up to second order in $J_{\perp,\parallel}/U$, the (again hard-core) model becomes $\hat H_{\text{eff}}=\hat H_0+\Delta \hat H$, where
\begin{eqnarray}
\Delta \hat H &=& -\frac{J_\perp^2}{U}\sum_{j} \hat n_{j,1}\hat n_{j,2}
- \frac{J_\parallel^2}{U}\sum_{j,\alpha} \hat n_{j,\alpha}\hat n_{j+1,\alpha}
\nonumber \\
&-&\frac{J_\perp J_\parallel}{2U}\sum_{j,\alpha,\beta\neq\alpha} 
\left [ 
\hat b_{j+1,\beta}^\dag \left (\hat n_{j+1,\alpha}+\hat n_{j,\beta} \right ) \hat b_{j,\alpha} +\mathrm{H.c.}
\right ] \nonumber\\
&-&\frac{J_\parallel^2}{2U}\sum_{j,\alpha}\left ( \hat b_{j+2,\alpha}^\dag \hat n_{j+1,\alpha}\hat b_{j,\alpha} + \mathrm{H.c.} \right ), 
\label{eq:Heff}
\end{eqnarray}
with nearest-neighbor~(NN) interactions along the legs and the rungs~(first line), and collisionally-assisted hops along plaquette diagonals~(second line), and between next-to-NN rungs~(third line).

\begin{figure}
\centering
\includegraphics[width=0.85\columnwidth]{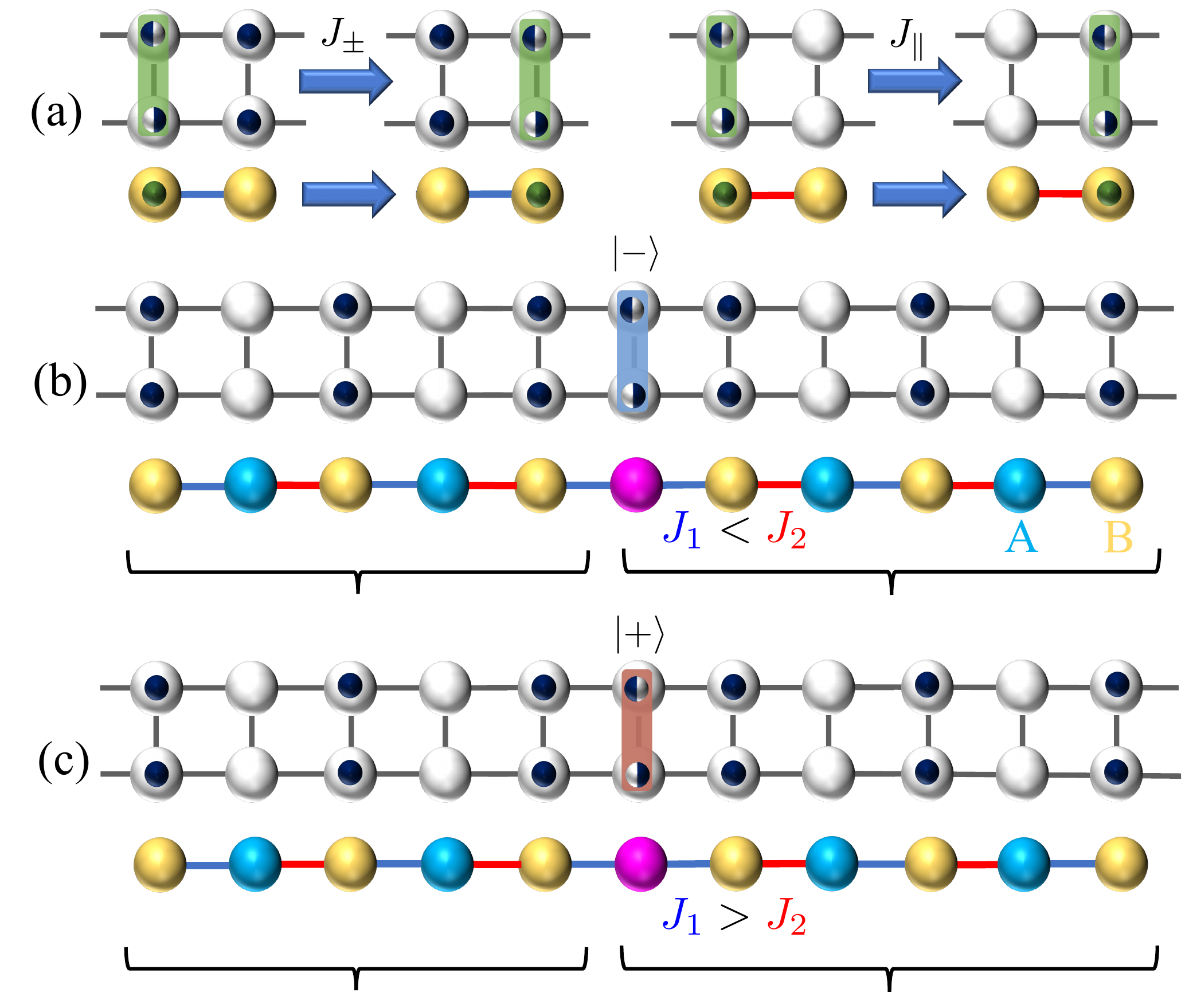}
\vspace{-0.3cm}
\caption{(a) A singlon defect propagates by swapping its position with doubly-occupied and empty rungs. Due to finite $U$, these swaps result in two different hopping rates~(blue and red lines). The defect propagates in an effective lattice in which the right and the left of the initial defect position present opposite SSH staggering, illustrated for the case of a defect in what should have been an empty rung. 
(b) For a $|-\rangle$ defect~(blue), $J_1<J_2$, and the defect experiences topological localization. (c) For a $|+\rangle$~(brown), $J_1>J_2$, and the defect displays non-topological localization. In the effective chain, light blue~(yellow) sites indicate the A~(B) sublattice.}
\label{fig:4}
\end{figure}

For $U=\infty$ leg-hopping induces the same swap rate for $|\pm\rangle$ with $|2\rangle$ or $|0\rangle$. For finite $U$, 
the interplay of leg hopping and collisionally-induced diagonal hopping causes $|\pm\rangle$ to swap with $|2\rangle$ at a rate 
$J_\pm = J_\parallel \left ( 1\pm 2J_\perp/U\right )$,
while $|\pm\rangle$ and $|0\rangle$ still swap at rate $J_\parallel$~(Fig.~\ref{fig:4}~(a)). This is particularly relevant if, as typically in experiments, the initial DW presents isolated singlon defects. 
Consider a singlon defect at rung 
$j=0$ in an otherwise perfect DW~(Figs.~\ref{fig:4}~(b-c)). 
The defect experiences an effective 
staggered hopping described by the SSH Hamiltonian:
\begin{equation}
H_{\mathrm{SSH}} = \frac{1}{2}\sum_{j} J_{j,j+1} \left (|\phi_j\rangle\langle \phi_{j+1}| +\mathrm{H. c.} \right ), 
\label{eq:HSSH}
\end{equation}
with $|\phi_j\rangle$ the state with the defect in rung $j$, and $J_{j,j+1}=J_1$~($J_2$) for even~(odd) $j$ for $j\geq0$, and the opposite for $j<0$.
The defect moves in an effective lattice of two SSH chains with opposite staggering that meet at the initial defect position~(Figs.~\ref{fig:4}~(b-c)). Since one SSH chain is topological and the other is not, an exponentially localized zero-energy mode appears at the interface~\cite{PhysRevLett.42.1698, Poli2015, Meier2016, Cai2019, Lin2021}.
We call the sublattice of even rungs
, where the defect is initially, A, and the sublattice of odd rungs B.



\paragraph{Topological and non-topological localization.--}
 We consider first $J_1<J_2$~(Fig.~\ref{fig:4}~(b)). 
This is the case in which a $|-\rangle$~($|+\rangle$) defect occurs where $|0\rangle$~($|2\rangle$) should have been, for which $J_1=J_-$
~($J_\parallel$) and $J_2=J_\parallel$~($J_+$). 
 At the interface between the SSH half-chains, the system presents a localized zero-energy edge mode~\cite{SM},  
$|\psi\rangle = \sqrt{P_0} \sum_m e^{-|m|/2\xi} |\phi_{2m}\rangle$,  
with $\xi = \frac{1}{2\ln(J_2/J_1)}$,
 and $P_0=|\langle \phi_0|\psi\rangle|^2 = (J_2^2-J_1^2)/(J_2^2+J_1^2)$. The defect~(initially in $|\phi_0\rangle$) 
remains hence localized with probability $P_0$, and the localized population only occupies A rungs. 



The situation changes radically if $J_1>J_2$, which corresponds to a $|-\rangle$~($|+\rangle$) defect where $|2\rangle$~($|0\rangle$) should have been, for which $J_1=J_\parallel$~($J_+$) and $J_2=J_-$~($J_\parallel$). 
Topological localization is precluded because the zero-energy edge state has only support in the B sublattice~\cite{SM}.
However, localized states appear at the two ends of the spectrum~\cite{SM}, 
with opposite energies $\pm E$, with 
$E\simeq \frac{1}{2}\left [
\sqrt{2} J_1 +\frac{J_2^2/4}{(\sqrt{2}-1)J_1}\right ]$. The population of these localized states, results in partial localization of the defect, which 
oscillates with frequency $\Omega=E/\hbar$ between the A and B sublattices~\cite{SM}. This may be intuitively understood from the case $J_2\ll J_1$, for which the central A site and the symmetric superposition of the neighboring B sites form an isolated two-level system characterized by an oscillatory frequency $\Omega=\sqrt{2}J_1/2\hbar$. 

\begin{figure}
    \centering
    \includegraphics[width=0.90\columnwidth]{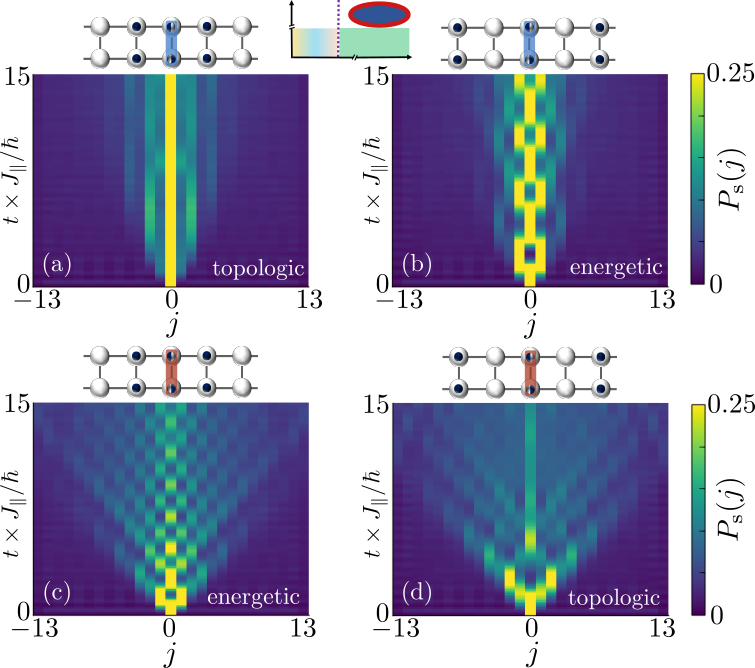}
    \vspace{-0.2cm}
    \caption{Time evolution of the probability of finding the defect in rung $j$, $P_{\mathrm{s}}(j)$. The color bar saturates above 0.25. The initial defect position and state is indicated on top of each panel, with a blue~(brown) box indicating a $\ket -$~($\ket +$) defect. The inset in the top locates this figure in Fig.~\ref{fig:1}(c).
    MPS simulations of the full Bose-Hubbard Hamiltonian $\hat H$ with 27 sites, $\eta = 16$, $J_\perp/U=1/2$, and bond dimension $\chi=768$.
    }
    \label{fig:5}
\end{figure}

Figure~\ref{fig:5} shows the 
time evolution of the singlon defect probability $P_s(j)$ for $\eta=16$ and $J_\perp/U=0.5$~\cite{BeyondPT}, for four different cases: topological localization in 
Figs.~\ref{fig:5}~(a) and~(d), and non-topological localization in Figs.~\ref{fig:5}~(b) and~(c).
Note the strikingly distinct dynamics in both cases, with clear oscillations between sublattice A and B in the non-topological case. Although the localization mechanism and the 
defect dynamics differ in the two regimes, the localized fraction is in any case very significant as long as $J_1$ and $J_2$ are sufficiently different. 
Hence, irrespective of the defect created and where it is produced, if $J_\perp/U$ is sufficiently large, we may expect strong localization of all singlon defects.


\paragraph{Other interaction-induced terms.--} The emergent SSH chain
experienced by the defects demands the 
stability of the DW
for finite  $U$ and $J_\parallel\ll J_\perp < U$.
Projecting $H_{\text{eff}}$ on the manifold of states without singlon defects~\cite{footnote-doublons}, we obtain an effective model for doubly-occupied rungs~\cite{SM}, 
$\hat H_D = \frac{3J_\parallel^2}{2U}\sum_j \left (\hat d_j^\dag \hat d_{j+1} + \hat d_{j+1}^\dag \hat d_j -\frac{10}{3}\hat n_{d,j} \hat n_{d,j+1} \right )$,
with $\hat d_j$ the (hard-core boson) annihilation operator of doubly-occupied rungs, and $\hat n_{d,j}=\hat d_j^\dag \hat d_j$. The hopping terms in $\hat H_D$
correspond to $\ket{20}\to \ket{02}$ swaps, which scramble the 
initial DW. Interestingly, the  relatively strong NN interaction between doubly-occupied rungs makes the DW stable against those swaps~\cite{SM}. Hence, quantum fluctuations of the DW order may lead for times $t>2U/3J_\parallel^2$ to the blurring of the background, but they do not destroy the localization~\cite{SM}.
Furthermore, 
 being hard-core bosons, singlon defects repel elastically~\cite{footnote-doublons}. NN interactions may lead to $|+,-\rangle\to |-,+\rangle$ swaps with a rate $J_\parallel^2/2U$, but these swaps are negligible for low defect densities~\cite{SM}. Finally,  collisionally-assisted next-to-NN hopping is only relevant due to hops $|\pm, 2, 0\rangle \to |0,2,\pm\rangle$, which scramble the DW. These processes occur with a rate $-J_\parallel^2/2U$, being neglible for $t<2U/J_\parallel^2$.




\paragraph{Experimental realization.--}
The emergent SSH chain experienced by each singlon defect is hence to a good approximation affected neither by the presence of neighboring defects, nor by quantum fluctuations of the DW or  next-to-NN hops. Probing the regime $J_\parallel\ll J_\perp<U$ is readily accessible in on-going experiments~\cite{wienand2023emergence}. 
Singlon-defect localization may be monitored in various ways. 
Current quantum gas microscopes~\cite{Impertro2024} can deterministically create and measure $\ket +$ and $\ket -$ with single-rung resolution. The dynamics depicted in Fig.~\ref{fig:5} can then be observed directly.
Even without site-resolved state preparation, asymmetries in the defect creation between even and odd rungs can be used to access the localization dynamics.
Under current experimental conditions~\cite{wienand2023emergence} we expect that $80\%$ of the created singlon defects are in even rungs, which should have been doubly occupied. One may hence monitor the 
imbalance ${\cal I}=2{\cal N}_E-1$, with ${\cal N}_E$ the number of singly-occupied even rungs. For $U=\infty$, defects propagate ballistically and ${\cal I}(t)={\cal I}(0){\cal J}_0(2J_\parallel t/\hbar)$. In contrast, topological and non-topological localization should result in a markedly different ${\cal I}(t)$, since topologically localized defects do not oscillate between sublattices, and non-topological defects oscillate between them with a frequency $\Omega$.

\paragraph{Conclusions.--} 
Bose-Hubbard ladders initialized in a density wave provide 
an unexpected platform for the realization of an emergent interaction-enabled topological-non-topological interface 
of two SSH chains with opposite staggering, without the necessity of tailored  external potentials. 
As a result, singlon defects experience two possible localization mechanisms, one topological and the other non-topological, characterized by strikingly distinct dynamics.
This surprising link between topology and many-body quantum dynamics can be probed in ongoing experiments. 
In a broader context, our work illustrates how defect dynamics is determined by the background substrate they move through (here a density wave). This idea readily generalizes to different static backgrounds, e.g. a random background may result in localization without external disorder akin to disorder-free localization \cite{PhysRevLett.118.266601}, and to dynamically coupled backgrounds, e.g. in bilayer settings defect motion reconfigures the initial state pattern and hence the effective tunnelling rates.
Our results also open further intriguing perspectives for the realization of emergent topological behavior in other platforms, in particular in synthetic ladders created using internal states~\cite{Mancini2015,Stuhl2015, Kolkowitz2017}.

{\it Acknowledgements} We acknowledge careful reading of the manuscript by Joanna Lis and Raphael Kaubruegger.
G.A.D.-C.~and L.S.~acknowledge the support of the Deutsche Forschungsgemeinschaft (DFG, German Research Foundation) -- Project-ID 274200144 -- SFB 1227 DQ-mat within the project A04, and under Germany's Excellence Strategy -- EXC-2123 Quantum-Frontiers -- 390837967). A.M.R.~and D.W.~acknowledge funding from AFOSR MURI Grant FA9550-21-1-0069, ARO  W911NF-24-1-0128,  NSF JILA-PFC PHY-2317149, OMA-2016244, the U.S. Department of Energy, Office of Science, National Quantum Information Science Research Centers, Quantum Systems Accelerator and  NIST.
M.A.~acknowledges funding from the Deutsche Forschungsgemeinschaft (DFG, German Research Foundation) via Research Unit FOR5522 under project number 499180199, via Research Unit FOR 2414 under project number 277974659 and under Germany’s Excellence Strategy – EXC-2111 – 390814868 and funding under the Horizon Europe programme HORIZON-CL4-2022-QUANTUM-02-SGA via the project 101113690 (PASQuanS2.1).

\bibliography{Topological}

\begin{thebibliography}{75}
\bibitem [{Bey()}]{BeyondPT}%
  \BibitemOpen
  \href@noop {} {}\bibinfo {note} {Note that $J_\perp/U=1/2$ lies clearly
  beyond the validity regime of perturbation theory, explaining why the
  localization in Fig.~4~(a) does not have zero localization length, as may
  have been expected from Eq. (5). However, the qualitative features of
  topological and non-topological localization remain valid.}\BibitemShut
  {Stop}%
\bibitem [{foo()}]{footnote-doublons}%
  \BibitemOpen
  \href@noop {} {}\bibinfo {note} {Similar to defect creation
  $|2,0\rangle\leftrightarrow|\pm,\pm\rangle$ is non-resonant due to the rung
  energy, and $|2,0\rangle\leftrightarrow|\pm,\mp\rangle$, which would be
  resonant, cannot be induced by leg hopping due to leg exchange
  symmetry.}\BibitemShut {Stop}%
\end{thebibliography}



\cleardoublepage
\appendix

\setcounter{equation}{0}
\setcounter{figure}{0}
\setcounter{table}{0}
\makeatletter
\renewcommand{\theequation}{S\arabic{equation}}
\renewcommand{\thefigure}{S\arabic{figure}}

%
\setcounter{equation}{0}
\setcounter{figure}{0}
\setcounter{table}{0}
\makeatletter
\renewcommand{\thefigure}{S\arabic{figure}}
\section{Supplementary Information}
This supplementary information contains additional details on the stability analysis of the DW in the hard-core regime, the defect theory explaining the staggered RRD correlations for $\eta>2$, the topological edge modes, the non-topologically localized states, the stability of the DW against swaps $20\to 02$, extended numerical results for lower $\eta$ values, details on the MPS simulations and convergence with time-step and bond-dimension, and defect-defect interactions.


\subsection{Stability analysis of the DW in the hard-core regime}
We briefly discuss the stability of the DW in the hard-core regime against the creation of singlon defects via Bogoliubov analysis valid at small excitation density (see also Refs.~\cite{bilitewski2023manipulating,dominguez2023relaxation} for similar analyses).

We introduce the hard-core bosonic operators $a_{i,\alpha}^\dag=b_{i,\alpha}$~($b_{i,\alpha}^\dag$) for even~(odd) $i$. The initial density wave is then mapped to a vacuum state, and the operators $a_{i,\alpha}^\dag$ create an excitation on top of the DW, i.e. a hole~(particle) in even~(odd) rungs. Note that this is also equivalent to a Holstein-Primakov transformation when mapping the hard-core Bose-Hubbard ladder as an XX spin ladder. Denoting 
$a_{i,\pm}=(a_{i,1}^\dag \pm a_{i,2}^\dag)/\sqrt{2}$, we re-write the Hamiltonian $\hat H_0$ 
in the form $\hat H_0 = \hat H_+ + \hat H_-$, where
\begin{eqnarray}
\!\!\!\!\hat H_\sigma \! =\!  -\!\sum_{i}\! \left [ 
\frac{\sigma  J_\perp}{2} \hat a_{i,\sigma}^\dag \hat a_{i,\sigma} 
\! -\! 
\frac{J_\parallel}{2} \! \left (
\hat a_{i,\sigma}^\dag \hat a_{i+1,\sigma}^\dag 
\! +\mathrm{H. c.} \! \right )\! \right ].
\end{eqnarray}
For a small density of excitations, we may neglect the hard-core character, and move to momentum space:
\begin{eqnarray}
\!\!\!\!\!\!\!\!\!\hat H_\sigma \! = \!  -\!\! \sum_{k}\!\! \left [ 
\!\frac{\sigma J_\perp}{2} \hat a_{k,\sigma}^\dag \hat a_{k,\sigma}
\!\! - \!
J_\parallel\!\cos k \! \left (
\hat a_{k,\sigma}^\dag \hat a_{-k,\sigma}^\dag 
\!\!+\!\mathrm{H. c.}\! \right )\! \right ],
\end{eqnarray}
which may be diagonalized by means of a Bogolioubov transformation, resulting in the 
dispersion law 
\begin{equation}
\xi(k)=\sqrt{J_\perp^2/4- 
J_\parallel^2 \cos^2 k}.
\end{equation}
Dynamical Bogoliubov instability is hence expected only 
for $J_\perp/J_\parallel<2$, as discussed in the main text.


\subsection{Defect theory explaining the staggered RDD correlations}
In this section we provide a non-interacting description of the dynamics in the two-defect sector on top of the initial DW pattern to explain the observed staggered RDD correlations. To do so we will first derive the dynamics of two non-interacting defects, and then in a second step explain how their propagation affects the RDD correlations.

Let us consider a pair of singlon defects, either $|+,+\rangle$ or $|-,-\rangle$, created by quantum fluctuations at neighboring rungs $j$ and $j+1$. Once created by quantum fluctuations, the defects may move along the ladder by swapping with $|2\rangle$ or $|0\rangle$, resulting~(for $U\to\infty$) in an effective hopping rate $J_\parallel$. Let us denote $|\phi_j\rangle$ as the 
state with a defect at rung $j$. The initial state of the pair is hence $|\phi_j,\phi_{j+1}\rangle$. We start by re-expressing this state in momentum space as
\begin{eqnarray}
|\phi_j,\phi_{j+1}\rangle = \frac{1}{L}\sum_{k,k'} e^{i (kj + k'(j+1))} |\tilde\phi_{k},\tilde\phi_{k'}\rangle.
\end{eqnarray}
Since $|\tilde\phi_k\rangle$ is an eigenstate of the leg hopping with eigenenergy $-J_\parallel\cos k$, 
the time-evolved state acquires the form
\begin{eqnarray}
&&\frac{1}{L}\sum_{k,k'} e^{i (kj + k'(j+1))} e^{iJ_\parallel t (\cos k + \cos k')} |\tilde\phi_{k},\tilde\phi_{k'}\rangle \nonumber \\
&=& \frac{1}{L^2}\sum_{k,k',l,l'} e^{i (kj + k'(j+1))} e^{iJ_\parallel t (\cos k + \cos k')} e^{-i(kl+k'l')} |\phi_l,\phi_{l'}\rangle \nonumber \\
&=& \sum_{l,l'} (-1)^j i^{1-l-l'} {\cal J}_{j-l}\left ( J_\parallel t \right ) {\cal J}_{j+1-l'}\left ( J_\parallel t \right ) |\phi_l,\phi_{l'}\rangle. 
\end{eqnarray}

We assume that the pair is initially created with equal probability at any position. Hence, initially, 
$|\psi(0)\rangle =\frac{1}{\sqrt{L}}\sum_j |\phi_{j},\phi_{j+1}\rangle$. Then, using that 
\begin{eqnarray}
&&\sum_{j,l,l'} (-1)^j i^{1-l-l'} {\cal J}_{j-l}\left ( J_\parallel t \right ) {\cal J}_{j+1-l'}\left ( J_\parallel t \right ) \nonumber \\ 
&=& \sum_{l,l'} i^{1+l-l'} {\cal J}_{l'-l-1} \left ( 2 J_\parallel t \right ),
\end{eqnarray}
we obtain the two-defect state at time $t$:
\begin{equation}
|\psi(t)\rangle = \frac{1}{\sqrt{L}}\sum_{l,l'\neq l} i^{1-(l'-l)} {\cal J}_{(l'-l)-1}(2J_\parallel t) |\phi_l,\phi_{l'}\rangle
\end{equation}
The probability to find a defect at rung $l$ and the other at rung $l'$ is then
\begin{equation}
P(l,l') = \frac{1}{L}\sum_{l,l'\neq l} \mathcal J_{(l'-l)-1}^2(2J_\parallel t) \, .
\end{equation}

Since this was derived for a pair-defect density of $1/L$, for the pair density of the order of $(J_\parallel/J_\perp)^2$, we obtain
\begin{equation}
P(l,l') \simeq  \left (\frac{J_\parallel}{J_\perp}\right )^2 \sum_{l,l'\neq l} {\cal J}_{(l'-l)-1}^2(2J_\parallel t).
\end{equation}
Hence, the number of defects at a distance $r$ is given by
\begin{equation}
P(r) =L \left (\frac{J_\parallel}{J_\perp}\right )^2 {\cal J}^2 _{|r|-1}(2J_\parallel t)
\end{equation}

We now turn to how these dynamics affect the background DW pattern and in turn the RDD correlations. Each time a defect moves it displaces a $|0\rangle$ or a $|2\rangle$. As a result, the region between the two defects consists of a string of sites with the DW pattern inverted compared to the initial state, i.e. a $|0\rangle$ becomes a $|2\rangle$, and vice versa (see Fig. 2(c) of the main text).  

In order to understand why this results in staggered RDD correlations, let us consider two rungs, for example 
$i=0$~(an even rung), and $j=r$, which may be even or odd. Although cumbersome, one can easily identify 
which changes occur in the occupations of those rungs, and with which probability they happen, when the pair of defects moves across those rungs. We employ in the following the notation (Change(i), Change(j)), where Change may be either "$=$"~(i.e. no change), $1$~(one particle in the rung), or Flip~(i.e. $|0\rangle\leftrightarrow |2\rangle$). 
These are the possible changes, and their probabilities:
\begin{itemize}
\item ($1$, $=$) or ($=$, $1$); $\sum_{l>0} P(l)+\sum_{0<l<r} P(l)$
\item  (Flip, $=$) or ($=$, Flip); $\sum_{l>0}\sum_{0<l'<r} P(l+l')$
\item (Flip, $1$) or ($1$, Flip); $\sum_{l>0} P(l+r)$
\item (Flip, Flip); $\sum_{l>0}\sum_{l'>0}P(l+l'+r)$
\item ($1$, $1$); $P(r)$
\end{itemize}
Each change affects the RDD correlations. After some tedious calculation one obtains:
\begin{eqnarray}
&&C_{0,r}(t)=(-1)^r \left [ 4\sum_ {j> 0}  j P(j+r, t) + P(r, t) \right ] \nonumber \\
&=& 2 (-1)^r \left ( \frac{J_\parallel}{{\cal J}_\perp} \right )^2 \left [ 4\sum_ {j> 0}  j {\cal J}_{j+r-1}(2J_\parallel t)^2 + {\cal J}_{r-1}(2J_\parallel t)^2 \right ] \nonumber \\
\end{eqnarray}
This expression corresponds to a ballistically-expanding cone of staggered correlations, which matches well with the numerical results of Fig. 2(d). 


\subsection{Topological edge modes and localisation}
In this section we provide the derivation for the topological edge modes appearing in the effective SSH model for single defect dynamics.

Let us consider the effective SSH model of Eq.~(4) of the main text. 
We can divide the lattice into A and B sublattice, such that $|m,A\rangle$~($|m,B\rangle$) is the A~(B) rung in the $m$-th elementary cell. The defect is initially in $|0,A\rangle$. We can then rewrite the Hamiltonian in the form:
\begin{eqnarray}
&&\hat H = - \frac{J_1}{2} \!\sum_{m\ge 0}\! \left [ |m,A\rangle \langle m,B| \!+\!   |m,B\rangle \langle m,A| \right ]
\nonumber \\
&&- \frac{J_2}{2} \!\sum_{m< 0}\! \left [ |m,A\rangle \langle m,B| \!+\!   |m,B\rangle \langle m,A|  \right ]
\nonumber \\
&&- \frac{J_2}{2} \!\sum_{m\ge 0}\! \left [ |m+1,A\rangle \langle m,B| \!+\!   |m,B\rangle \langle m+1,A| \right ]
\nonumber \\
&&- \frac{J_1}{2} \! \sum_{m< 0}\! \left [ |m+1,A\rangle \langle m,B| \!+\!   |m,B\rangle \langle m+1,A| \right ].
\end{eqnarray}


\begin{figure}[t!]
    \centering
    \includegraphics[width=\columnwidth]{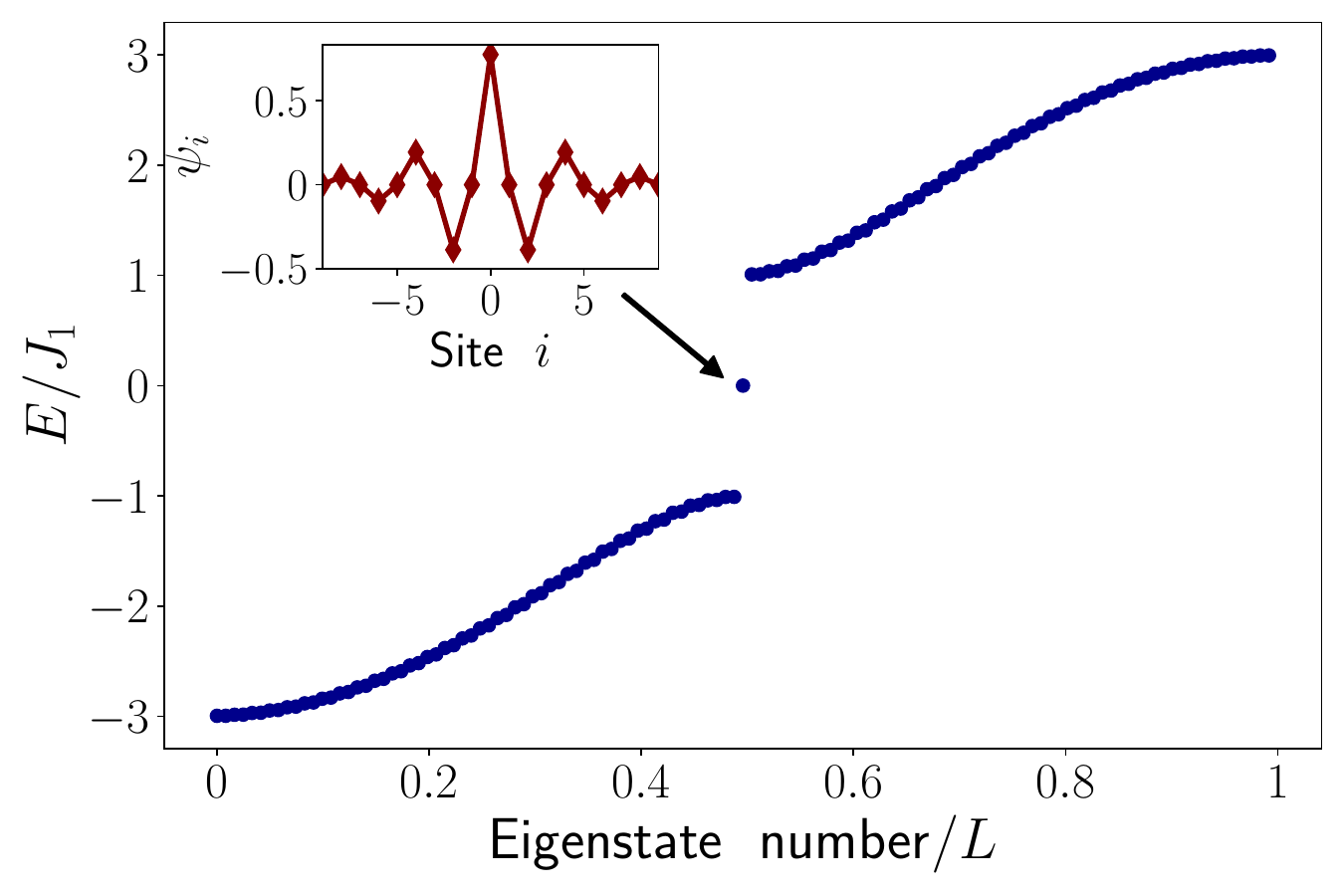}
    \caption{Eigenspectrum of the SSH model of Eq.~(4) of the main text, when $J_2=2J_1$. The inset shows the wavefunction $\psi_i$ at site $i$~(with $i=0$ the initial defect position) of the topological zero-energy edge mode. Note that the mode occupies only sublattice A~(even sites).\label{fig:S1}}
\end{figure}


We consider first the case $J_1<J_2$. As shown in Fig.~\ref{fig:S1} 
the eigenspectrum of this model presents a topological zero-energy mode at the interface between the two SSH chains. In order to understand the properties of this edge mode, let us consider a state $|\psi\rangle = \sum_m \left [a_m|m,A\rangle + b_m |m,B\rangle \right ]$. If this state is the topological edge state with eigenstate $E=0$, then $\hat H |\psi\rangle=0$, which leads to the 
recursions:
\begin{itemize}
\item For $m\geq 0$, $a_{m+1}=-\frac{J_1}{J_2}a_m$
\item For $m< 0$, $a_{m-1}=-\frac{J_1}{J_2}a_m$
\item $b_0+b_{-1}=0$
\item For $m\geq 0$: $b_{m+1}=-\frac{J_2}{J_1}b_m$
\item For $m< 0$: $b_{m-1}=-\frac{J_2}{J_1}b_m$ 
\end{itemize}
Then: $a_m = \left (-\frac{J_1}{J_2} \right )^{|m|} a_0$, and 
$b_{m\geq 0} = \left (-\frac{J_2}{J_1} \right )^{m} b_0 = -b_{-(m+1)}$. 
If $J_1/J_2<1$, we can define $J_1/J_2 \equiv e^{-1/2\xi}$, such that 
$|a_m|^2 = e^{-|m|/\xi} |a_0|^2$ and 
$|b_{m\geq 0}|^2 = e^{m/\xi} |b_0|^2 = |b_{-(m+1)}|^2$.
Since the solution must be normalized, this means that the physical solution must have 
$b_0=0$, and hence the localized edge mode lives only in the A sublattice:
$|\psi\rangle = a_0 \sum_m e^{-|m|/2\xi} |m,A\rangle$. We illustrate this localised state in the inset of Fig.~\ref{fig:S1}.

To connect with the expansion dynamics of an initially localised defect, starting with $|0,A\rangle$ we expect that a probability given by $P_0=|a_0|^2$ remains localized in the edge state, which lives only in the A sublattice. The value of the localized fraction $P_0$ is easily 
found by normalization:
\begin{equation}
P_0 = \frac{J_2^2-J_1^2}{J_2^2+J_1^2}.
\end{equation}


\subsection{Non topological localized states}
Next we consider $J_1>J_2$, in which a defect on the A sublattice turns out to be non-topologically localised.

\begin{figure}[t!]
    \centering
    \includegraphics[width=\columnwidth]{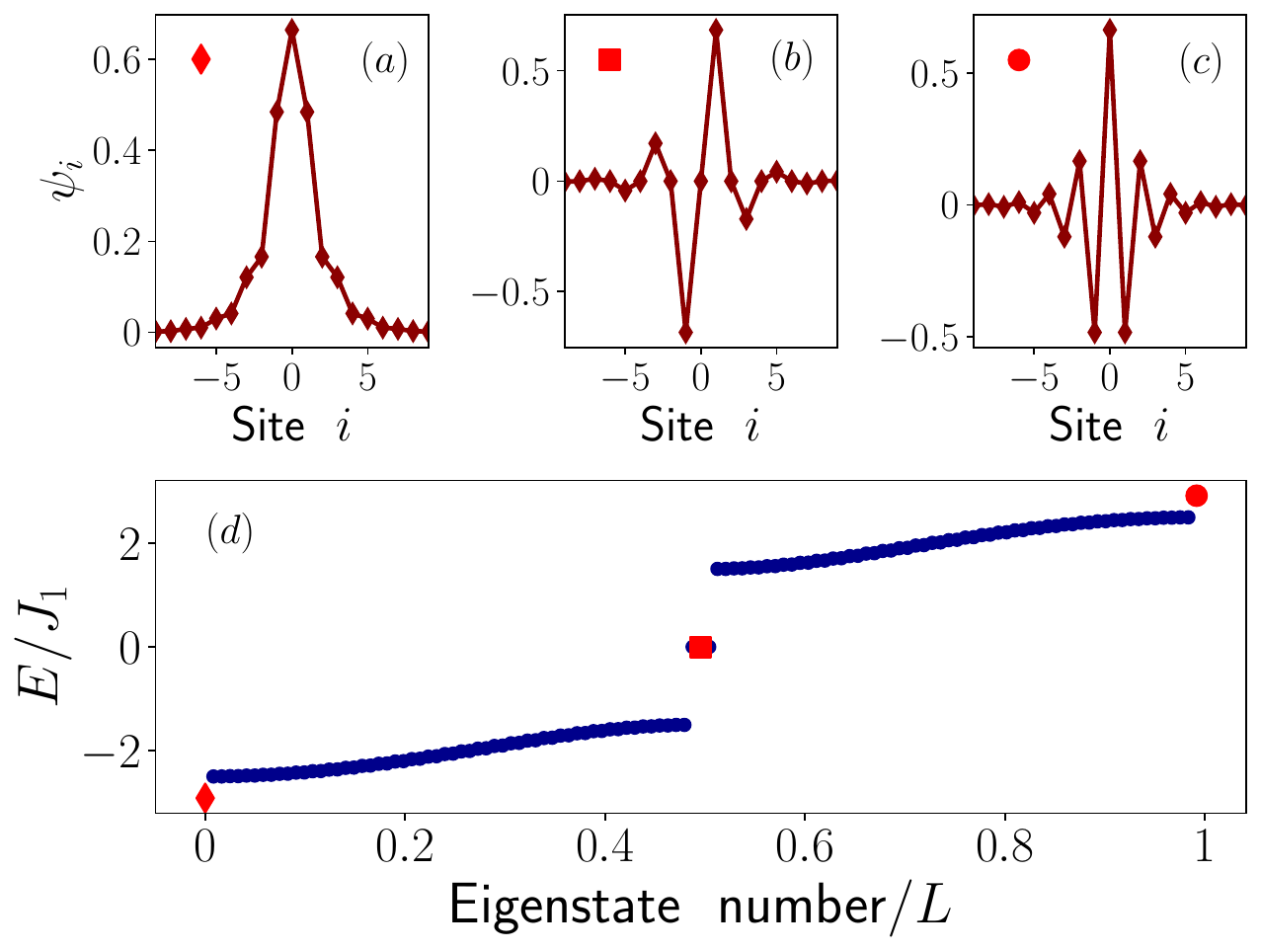}
    \caption{(a-c) Wavefunction 
    $\psi_i$ at site $i$~(with $i=0$ the initial defect position) of the localized eigenstates of model~(4) of the main text, when $J_1=2J_2$. 
    (d) Eigenspectrum of the model, where we have indicated with different symbols the energies of the localized states depicted in panels (a-c). The zero-energy mode only exists in sublattice B (odd rungs). The other localized solutions correspond to eigenstates at both ends of the spectrum. Note that $\psi_i$ in panel (a) corresponds to $(-1)^i \psi_i$ in panel (c).}
    \label{fig:S2}
\end{figure}

The corresponding eigenspectrum is depicted in Fig.~\ref{fig:S2}~(d). The same analysis as in the previous section shows that for $J_1>J_2$ the zero-energy topological edge mode only occupies sublattice B, see Fig.~\ref{fig:S2}~(b), and, thus, is not relevant to the dynamics of a defect initially on the A sublattice. However two other localized modes, which have overlap with the central site, appear at both ends of the eigenspectrum, see Fig.~\ref{fig:S2} (a) and (c). 

In order to understand the properties of those localized states, we define $|\alpha_0\rangle \equiv |0,A\rangle$, $|\alpha_{m>0}\rangle \equiv \frac{|m,A\rangle + |-m,A\rangle}{\sqrt{2}}$, and 
$|\beta_{m\geq 0}\rangle \equiv \frac{|m,B\rangle + |-(m+1),B\rangle}{\sqrt{2}}$. 
Let us consider states $|\psi\rangle = \sum_{m\geq 0} \left ( \alpha_m |\alpha_m \rangle +  \beta_m |\beta_m \rangle\right )$. For the states of interest, i.e. the manifold of states that may be reached from $|0,A\rangle$, 
we can define the Hamiltonian:
\begin{eqnarray}
\hat H &=& -\sqrt{2} \frac{J_1}{2} \left ( |\alpha_0\rangle\langle \beta_0 | + \mathrm{H.c.}\right ) \nonumber \\
&-& \frac{J_2}{2} \sum_{m\geq 0} \left (|\beta_m\rangle\langle \alpha_{m+1}| + \mathrm{H.c.}\right ) \nonumber \\
&-& \frac{J_1}{2} \sum_{m>0} \left (|\alpha_m\rangle\langle \beta_m| + \mathrm{H.c.}\right ).
\end{eqnarray}
Let $|S_m\rangle \equiv \frac{|\alpha_m\rangle + |\beta_m\rangle}{\sqrt{2}}$, and 
$|A_m\rangle \equiv \frac{|\alpha_m\rangle - |\beta_m\rangle}{\sqrt{2}}$, then:
\begin{eqnarray}
&& \hat H = -\sqrt{2} \frac{J_1}{2} \left ( |S_0\rangle\langle S_0 | - |A_0\rangle\langle A_0 |\right ) \nonumber \\
&-& \frac{J_1}{2} \sum_{m>0} \left ( |S_m\rangle\langle S_m | - |A_m\rangle\langle A_m |\right )
\nonumber \\
&-& \frac{J_2}{4} \sum_{m\geq 0} \left (|S_m\rangle\langle S_{m+1}| - |A_m\rangle\langle A_{m+1}|
\right \delimiter 0 \nonumber \\
&&+ \left\delimiter 0 |S_m\rangle\langle A_{m+1}| - |A_m\rangle\langle S_{m+1}| +  \mathrm{H.c.}\right ).
\label{eq:HSA}
\end{eqnarray}


Assuming $J_2 /J_1 \ll 2(\sqrt{2}-1)$, we can neglect the coupling between $A$ and $S$ states, and hence 
obtain $\hat H\simeq \hat H_S+\hat H_A$, with 
\begin{eqnarray}
\hat H_S &=& -\sqrt{2} \frac{J_1}{2} |S_0\rangle\langle S_0 |
-\frac{J_1}{2} \sum_{m>0} |S_m\rangle\langle S_m | \nonumber \\
&-& \frac{J_2}{4} \sum_{m\geq 0} \left |S_m\rangle\langle S_{m+1}| +  \mathrm{H.c.}\right ).
\end{eqnarray}
and $\hat H_A$ is the same with an overall minus. We can remove the identity $-J_1 \mathbf{1}$, obtaining the final expression
\begin{eqnarray}
\hat H_S &=& -(\sqrt{2}-1) \frac{J_1}{2} |S_0\rangle\langle S_0 | \nonumber \\
&-& \frac{J_2}{4} \sum_{m\geq 0} \left ( |S_m\rangle\langle S_{m+1}| +  \mathrm{H.c.}\right ),
\end{eqnarray}
which describes  a semi-infinite lattice with a defect in the first site.

Let us assume an exponentially localized solution $|\psi_S\rangle = {\cal N} \sum_{m\geq 0} e^{-m/2\xi} |S_m\rangle$, 
where the normalization ${\cal N}^2$ is the probability to be in $|S_0\rangle$.
This solution has an energy
$E=-\sqrt{2}\frac{J_1}{2} -\frac{J_2}{4}e^{-1/2\xi}$, where we restored the $-J_1$ we had dropped. 
On the other hand, second-order perturbation on top of the solution $|S_0\rangle$, gives the energy
\begin{eqnarray}
E\simeq -\sqrt{2}\frac{J_1}{2} -\frac{J_2^2/2}{(\sqrt{2}-1)J_1}. 
\end{eqnarray}
We can hence approximate
$e^{-1/2\xi}\simeq \frac{J_2/J_1}{2({\sqrt{2}-1})}$. 
Since ${\cal N}^2 = 1-e^{-1/\xi}$, we obtain
\begin{equation}
{\cal N}^2 = 1-\left ( \frac{J_2/J_1}{2(\sqrt{2}-1)}\right )^2.
\end{equation}
This then fully determines the localized states discussed in the main text. 

Note that the localized states at both ends are such that 
the state $\psi_i^{(E<0)}$ with energy $E<0$ and the states $\psi_i{(E>0)}$ with energy $E>0$ fulfill $\psi_i^{(E<0)}=(-1)^i\psi_i^{(E>0)}$. As a result a linear combination $\psi_i^{(E<0)}+\psi_i^{(E>0)}$ oscillates with frequency $\Omega=E/\hbar$ between the A and the B sublattice, as described in the main text.


\subsection{Stability of the DW against \texorpdfstring{$20 \to 02$}{20 <-> 02} swaps}

We project the second-order Hamiltonian $\hat H_{\mathrm{eff}}$, see Eq.~\eqref{eq:H0} and Eq.~\eqref{eq:Heff}, into the manifold without singlon defects (doublon-holon manifold).

 The case of neighboring doubly-occupied rungs~(doublons) is simple since $\hat H_{\mathrm{eff}}$ does 
 act diagonally on that state
\begin{equation}
\hat H_{\mathrm{eff}}|2,2\rangle = -\frac{2J_\parallel^2}{U}|2,2\rangle,
\end{equation}
where we disregard the rung energy because it will be a constant in the doublon-holon manifold.

Let us consider now a neighboring doublon-holon pair, i.e. $|2,0\rangle$. 
Let us split the $\hat H_{\mathrm{eff}}= \hat H_{\mathrm{eff},0} + \hat H_{\mathrm{eff},1}$, where 
\begin{eqnarray}
\!\!\!\!\!\!\!\!\!\!\!\!\!\!\hat H_{\mathrm{eff},0} \!=\! -\frac{J_\perp}{2}\! \sum_{j} \left (\hat b_{j,1}^\dag \hat b_{j,2} + \mathrm{H.c}\right )
\!\!-\frac{J_\perp^2}{U}\sum_{j} \hat  n_{j,1} \hat n_{j,2},
\label{eq:Heff0}
\end{eqnarray}
is the rung energy. Note that $\hat H_{\mathrm{eff},0}|2\rangle=-\frac{J_\perp^2}{U}|2\rangle$, 
$\hat H_{\mathrm{eff},0}|\pm\rangle = \mp \frac{J_\perp}{2} |\pm\rangle$, and 
$\hat H_{\mathrm{eff},0}|0\rangle=0$. We consider $\hat H_{\mathrm{eff},1}$ as a perturbation to $\hat H_{\mathrm{eff},0}$. 
Then:
\begin{eqnarray}
\hat H_{\mathrm{eff},1}|\pm,\pm\rangle &=& -\frac{J_\parallel^2}{2U}\left ( |+,+\rangle + |-,-\rangle \right )
\nonumber \\
&\mp&  \frac{J_\parallel}{2} \left ( 1\pm \frac{J_\perp}{U}\right) \left ( |2,0\rangle + |0,2\rangle \right )
\\
\hat H_{\mathrm{eff},1}|2,0\rangle &=& -\frac{J_\parallel}{2}\left (1+\frac{J_\perp}{U} \right ) |+,+\rangle
\nonumber \\
&+&\frac{J_\parallel}{2}\left (1-\frac{J_\perp}{U} \right ) |-,-\rangle. 
\label{eq:Heff_20}
\end{eqnarray}
We can then easily evaluate the second-order processes:
\begin{eqnarray}
&&\hat H^{(2)}|2,0\rangle 
\nonumber \\
&=&
\left [\frac{J_\parallel^2 (1+J_\perp/U)^2}{4(J_\perp-J_\perp^2/U)}
-\frac{J_\parallel^2 (1-J_\perp/U)^2}{4(J_\perp+J_\perp^2/U)} \right ] \left ( |2,0\rangle + |0,2\rangle \right ) 
\nonumber \\
&\simeq& \frac{3J_\parallel^2}{2U} \left ( |2,0\rangle + |0,2\rangle \right ),
\end{eqnarray}
where we assume $J_\perp/U\ll 1$. 
Note that although the result for $\hat H^{(2)}|2,0\rangle$ only depends on $J_\parallel^2/U$,  
it stems from assuming that $J_\perp$ is much larger than any energy that can 
take out from the doublon-holon manifold.


We can then write the Hamiltonian for the manifold with only doublons and holons in the following way:
\begin{eqnarray}
\hat H_{D} &=& \frac{3J_\parallel^2}{2U}
\sum_j \left [ (\hat d_j^\dag \hat h_j)(\hat h_{j+1}^\dag \hat d_{j+1})+(\hat h_j^\dag \hat d_j)(\hat d_{j+1}^\dag \hat h_{j+1}) \right ]
\nonumber \\
&-& \frac{2J_\parallel^2}{U}\sum_{j} \hat n_{d,j} \hat n_{d,j+1} 
\nonumber \\
&+& \frac{3J_\parallel^2}{2U}\sum_j
\left (\hat n_{d,j} \hat n_{h,j+1} + \hat n_{h,j} \hat n_{d,j+1}\right ),
\end{eqnarray}
where $\hat d_j$~($\hat h_j$) destroys a doublon~(holon) at rung $j$, $\hat n_{d,j}=\hat d_j^\dag \hat d_j$, $\hat n_{h,j}=\hat h_j^\dag \hat h_j$, and $\hat n_{d,j}+\hat n_{h,j}=1$.  Note that there 
is no term arising from the collisionally-assisted next-to-NN hops, since that term would lead to the destruction of doublons and holons, which 
would take the system out the doublon-holon manifold. 
Since, destroying~(creating) a doublon implies creating~(destroying) a holon, 
since $\hat n_{h,j}=1-\hat n_{d,j}$, and since the number of holons and doublons is conserved, 
we may rewrite:
\begin{equation}
\hat H_{D} = \frac{3J_\parallel^2}{2U}\sum_j \left (\hat d_j^\dag \hat d_{j+1} + \hat d_{j+1}^\dag \hat d_j -\frac{10}{3}\hat n_{d,j} \hat n_{d,j+1} \right ),
\end{equation}
which is the expression written in the main text.

The relatively strong NN interaction between doubly-occupied sites is crucial for the DW stability. Considering an initial perfect DW with doubly-occupied even rungs, we introduce the operators that characterize alterations~(DW-defects) of the initial DW order (i.e. a $|2\rangle$~($|0\rangle$) where there should be a $|0\rangle$~($|2\rangle$)):  $\hat c_j=\hat d_j^\dag$~($\hat d_j$) and 
$\hat n_{c,j}=1-\hat n_{d,j}$~($\hat n_{d,j}$) for even~(odd) $j$. 
This transforms the Hamiltonian into:
\begin{eqnarray}
\hat H_{DH} &=& \frac{3J_\parallel^2}{2U}
\sum_j \left ( \hat c_j^\dag \hat c_{j+1}^\dag 
+ \hat c_j \hat c_{j+1} 
\right\delimiter 0 \nonumber \\
&+&\left\delimiter 0 
\frac{10}{3} (1/2-\hat n_{c,j})(1/2-\hat n_{c,j+1}) \right ).
\end{eqnarray}
Assuming a dilute gas of DW-defects, we may neglect the 
defect-defect interaction, and removing constants we get:
\begin{eqnarray}
\hat H_{DH} &=& \frac{3J_\parallel^2}{2U}
\sum_j \left ( \hat c_j^\dag \hat c_{j+1}^\dag + \hat c_j \hat c_{j+1} +\frac{10}{3} \hat c_j^\dag \hat c_j \right ).
\end{eqnarray}
Moving to momentum representation:
\begin{eqnarray}
\hat H &=& \frac{3J_\parallel^2}{2U}
\sum_{k>0} \left [ 2\cos(k) \left (\hat c_k^\dag \hat c_{-k}^\dag + \hat c_k \hat c_{-k} \right )
\right\delimiter 0\nonumber\\
&+&\left\delimiter 0 \frac{10}{3} \left (
\hat c_k^\dag \hat c_k +\hat c_{-k}^\dag \hat c_{-k}\right )\right ].
\end{eqnarray}
Let $\hat \beta_k = u_k \hat c_k -v_k \hat c_{-k}^\dag$, we impose $\xi \hat \beta_k = [\hat H_{DH},\hat \beta_k]$, which leads to the Bogoliubov-de Gennes equations:
\begin{eqnarray}
\xi(k) u_k = \frac{3J_\parallel^2}{2U} \left [ \frac{10}{3}u_k - 2\cos(k) v_k\right ], \nonumber \\
\xi(k) v_k = \frac{3J_\parallel^2}{2U} \left [ 2\cos(k) u_k -\frac{10}{3}v_k \right ], \nonumber
\end{eqnarray}
Diagonalizing the Bogoliubov-de Gennes matrix, we obtain the dispersion:
\begin{equation}
\xi(k)=\frac{3J_\parallel^2}{U}\sqrt{\left ( \frac{5}{3}\right )^2-\cos^2k}.
\end{equation}
Therefore, irrespective of $k$, the dispersion remains real, and hence the DW is Bogoliubov 
stable against the formation of DW-defects.



\begin{figure}[t]
    \centering
\includegraphics[width=\columnwidth]{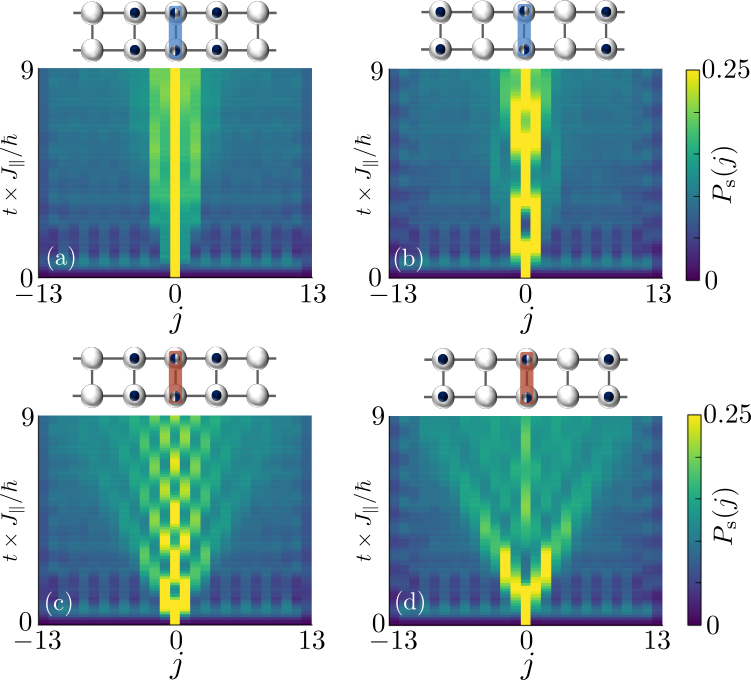}
    \caption{Defect motion for $\eta = 8$, $U / J_\parallel = 16$, otherwise same as Fig.~\ref{fig:5} of the main text. When reducing $\eta$, perturbative creation of defects is less suppressed. As a consequence, a background of singlon defects forms, but despite this background localization remains clearly visible.}
\label{fig:background_pair_creation}
\end{figure}


\subsection{Singlon-defect localization despite fluctuations of the DW pattern}
In this section we address the robustness of the described phenomenology of topological and non-topological localisation obtained within the single defect effective SSH model to relevant fluctuations of the background pattern.

As mentioned in the main text, although the DW is stable against the creation of singlon defects, a small density of defects of the order of $1/\eta^2$ is created by quantum fluctuations, which creates pairs $|+,+\rangle$ and $|-,-\rangle$. As the ratio $\eta$ decreases, a growing background of quantum defects develops. Fig.~\ref{fig:background_pair_creation} shows that even in the presence of a small fraction of quantum singlon-defect pairs, localization remains stable at experimentally relevant times for $\eta = 8$ and $U/J_\parallel=16$.

\subsection{MPS simulations}
We now describe the matrix product state simulation of the full Bose-Hubbard model.
We represent the ladder as a one-dimensional matrix product state (MPS) by snaking through the ladder as $(i, \alpha) \rightarrow s=2i+\alpha$. 
Since the bosons are strongly interacting, we truncate the bosonic Hilbert space to at most two bosons per lattice site.
We use the time-evolving block decimation (TEBD) to compute time evolution of the state~\cite{vidal2004efficient}. This approach is based on decomposing the Hamiltonian into two-site gates acting on nearest neighbors. To include tunneling along the ladder legs, we include swap gates along the rungs. We further include on-site interactions and rung tunneling in the same two-site Hamiltonian to reduce Trotter errors. We use a 4th order Trotter decomposition~\cite{sornborger1999higher}, allowing reasonably-sized time steps even in the presence of fast on-site interactions. In particular, we use the decomposition
\begin{widetext}
\begin{align}
    (1)^T(1)(1)^T(-2)(1)^T(1)^T(1)^T(1)^T(1)(1)^T(1)(1)(1)(1)(-2)^T(1)(1)^T(1) \, .
\end{align}
\end{widetext}
Here, $(1)^T$ signifies applying the gates from the leftmost site of the MPS to the rightmost site, $(1)$ signifies the reverse order, and $(-2)$/$(-2)^T$ are Trotter steps with a time step $-2dt$. Our simulation enforces the conservation of the number of particles. It is written in the julia programming language~\cite{bezanson2017julia} and builds on the ITensor package~\cite{fishman2022itensor}.


\begin{figure}[t]
    \centering
    \includegraphics[width=\columnwidth]{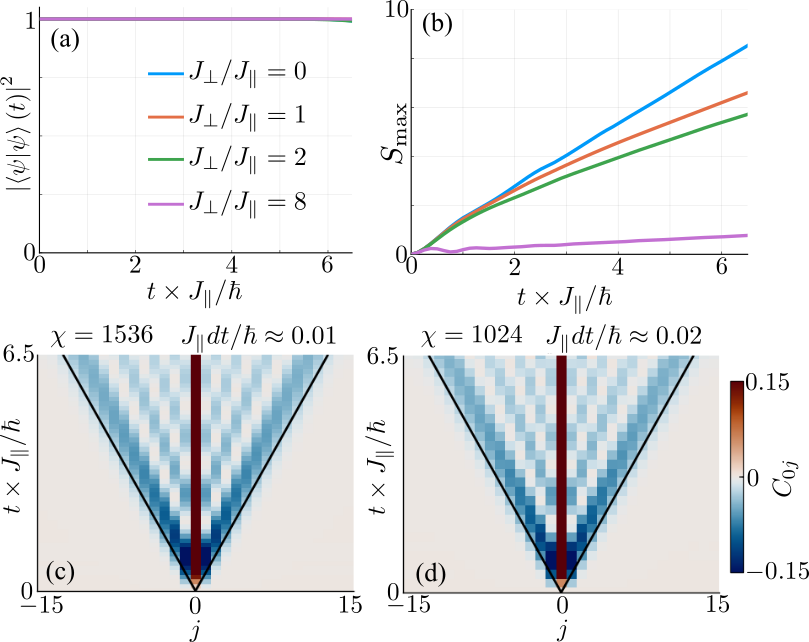}
    \caption{Convergence of MPS data corresponding to Fig.~2.
    (a) Square of the state norm for the different ratios of $J_\perp/J_\parallel$. The curves overlap almost perfectly.
    (b) Maximal bipartite entanglement entropy for any MPS bipartition for different rations of $J_\perp / J_\parallel$.
    (c/d) Demonstration of convergence by plotting rung-rung correlations for two different bond dimensions in the worst-case scenario (largest entropy).
    }
    \label{fig:convergence-2}
\end{figure}



\begin{figure}[t]
    \centering
    \includegraphics[width=\columnwidth]{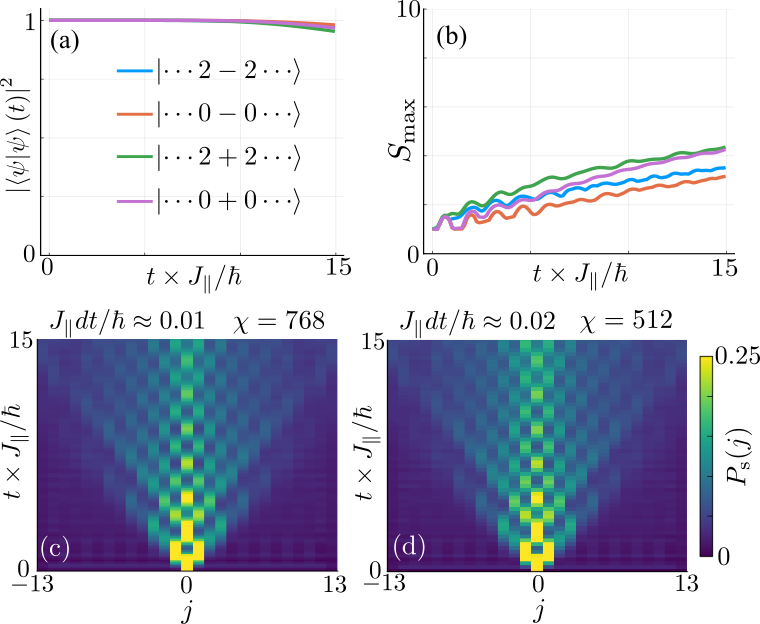}
    \caption{Convergence of MPS data corresponding to Fig.~4.
    (a) Square of the state norm for the different initial defects.
    (b) Half-chain entanglement entropy for the different initial defects. The cut along which the entanglement entropy is considered is shown as a dotted line in the inset.
    (c/d) Demonstration of convergence by plotting rung-rung correlations for two different time steps in the worst-case scenario (largest entropy).
    }
    \label{fig:convergence-4}
\end{figure}


To test convergence with both time step and bond dimension, we repeat the simulations with a smaller bond dimension and larger time step and confirm that all plots remain virtually indistinguishable. In particular, we change the time step between $J_\parallel dt / \hbar = 2\pi \times 0.02/12$ and $J_\parallel dt / \hbar = 2\pi \times 0.04/12$, and iteratively change the bond dimension as $\chi = 128, 256, 512, 768, 1024, 1536$. The results of this procedure are shown in Fig.~\ref{fig:convergence-2}(c/d) for the data corresponding to Fig.~\ref{fig:cones} of the main text, and in Fig.~\ref{fig:convergence-4}(c/d) for the data corresponding to Fig.~\ref{fig:5}. Other than blurring out due to worse time resolution, there is no visible change between the plots.

We further test convergence with the bond dimension $\chi$ via the state norm and the entropy. For each applied gate, TEBD performs a singular value decomposition and truncates the smallest singular values after a given bond dimension. This procedure reduces the state norm by the square of the truncated singular values. Therefore, as long as the norm remains close to one, the state is barely truncated, and the MPS is likely a faithful representation of the true state. The state norm as a function of time is shown in Fig.~\ref{fig:convergence-2}(a) and Fig.~\ref{fig:convergence-4}(a) and remains well above 0.9 in all cases, indicating that the results are indeed converged. As a leading order corrections, whenever we compute observables, we divide the result by the state norm. In addition, we keep track of the maximal entanglement entropy for each bipartition of the chain throughout the evolution. The entropy is upper-bounded by the bond dimension $\chi$ as $S_\mathrm{max} \leq \log_2(\chi)$, and for convergence one typically wants $\chi \gg 2^{S_\mathrm{max}}$. This is the case for all parameters analyzed. Finally, we also confirmed the convergence of the entanglement entropy with bond dimension and found that it changes by at most 0.1 for all cases when decreasing the bond dimension in the above-mentioned steps (not shown). This indicates that all results are fully converged.

\subsection{Defect-defect interactions}


\begin{figure}[hbtp]
    \centering
    \includegraphics[width=\columnwidth]{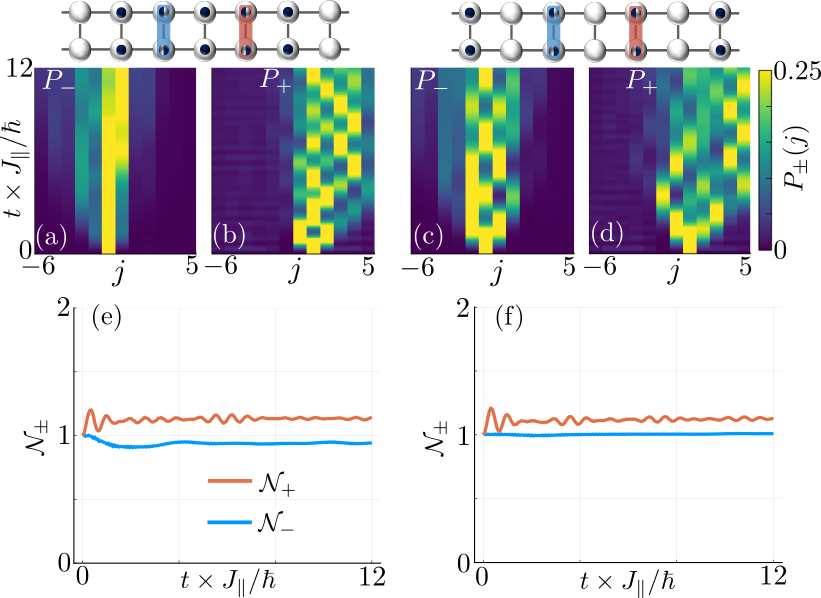}
    \caption{Collision of two defects. (a)-(d) Probabilities to find each singlon defect in the different rungs for an initial singlon defect arrangement shown in the top row (identical for (a) and (b), and (c) and (d), respectively.
    (e) Cumulative number of singlon defects in the entire chain for the arrangement in (a/b).
    (f) Cumulative number of singlon defects in the entire chain for the arrangement in (c/d).
    Parameters: $\eta = 16$, $U/J_\parallel = 32$, bond dimension $\chi = 512$, chain length $L=12$.
    }
    \label{fig:2defects}
\end{figure}


In the effective model, all singlon defect collisions are elastic and thus leave the physics discussed in the paper largely unperturbed. While inelastic $\ket{+,+}$ and $\ket{-,-}$ collisions are energetically suppressed, inelastic $\ket{-,+}$ transitions are forbidden by leg exchange symmetry. The only inelastic process is the perturbatively slow swap $\ket{-,+} \leftrightarrow \ket{+,-}$ at rate $J_\parallel^2/2U$.

In Fig.~\ref{fig:2defects}, we numerically verify that indeed $\ket{-,+}$ collisions do not significantly alter the conclusions drawn in the paper, such that they will be observable even in the presence of background defects.
We initialize a $\ket -$ and a $\ket +$ defect separated by one site as shown in the top row. We then consider the time evolution of the probability to find a $\ket -$ or $\ket +$ defect independently in (a,b) and (c,d) for each initial configuration. In both cases, the $\ket -$ defect remains in the left half of the system, while the $\ket +$ defect remains in the right half, indicating that indeed swapping of the two defects is slow compared to the time scales of interest. Furthermore, the $\ket -$ state remains strongly localized in both cases despite the presence of a second defect.

Panels (e) and (f) show the total number of $\ket -$ and $\ket +$ defects in the chain. As predicted by the perturbative calculation, pair-annihilation of $\ket -$ and $\ket +$ defects is very inefficient since both $\cal N_+$ and $\cal N_-$ drop barely below one. These results also show a stark difference in the number of $\ket +$ and $\ket -$ defects which are dynamically created from the background. This is explained by a difference in the effective creation rates Eq.~\eqref{eq:Heff_20}, and the smaller energetic cost for creating $\ket +$ defects due to the superexchange energy of the initial state $\ket{2,0}$. 

\end{document}